\documentclass[%
 reprint,
superscriptaddress,
 amsmath,amssymb,
 aps,
 prc
]{revtex4-1}

\usepackage{graphicx}
\usepackage{dcolumn}
\usepackage{bm}
\usepackage{xfrac}
\usepackage{mathtools}
\usepackage{amsmath}
\usepackage{xcolor}

\begin{document}

\preprint{APS/123-QED}

\title{Proton decays from $\alpha$-unbound states in $^{22}$Mg and the $^{18}$Ne($\alpha,p_0$)$^{21}$Na cross section}

\author{J. W. Br\"{u}mmer}
\email{jwbrummer@tlabs.ac.za}
\affiliation{Department of Physics, Stellenbosch University,
Private Bag X1, 7602 Matieland, Stellenbosch, South Africa}
\affiliation{iThemba Laboratory for Accelerator Based Sciences, Somerset West 7129, South Africa}

\author{P. Adsley}%
\affiliation{iThemba Laboratory for Accelerator Based Sciences, Somerset West 7129, South Africa}
\affiliation{School of Physics, University of the Witwatersrand, Johannesburg 2050, South Africa}
\affiliation{Cyclotron Institute , Texas A\&M University, College Station, Texas 77843-3366, USA}
\affiliation{Department of Physics \& Astronomy, Texas A\&M University, College Station, Texas 77843-4242, USA}

\author{T. Rauscher}
\affiliation{Department of Physics, University of Basel, Basel, Switzerland}
\affiliation{Centre for Astrophysics Research, University of Hertfordshire, Hatfield AL10 9AB, United Kingdom}

\author{F. D. Smit}
\affiliation{iThemba Laboratory for Accelerator Based Sciences, Somerset West 7129, South Africa}

\author{C. P. Brits}
\affiliation{Department of Physics, Stellenbosch University,
Private Bag X1, 7602 Matieland, Stellenbosch, South Africa}

\author{M. K\"{o}hne}
\affiliation{Department of Physics, Stellenbosch University,
Private Bag X1, 7602 Matieland, Stellenbosch, South Africa}

\author{N. A. Khumalo}
\affiliation{Department of Physics, University of the Western Cape, Private Bag X17, Bellville 7535, South Africa}

\author{K. C. W. Li}
\affiliation{Department of Physics, Stellenbosch University,
Private Bag X1, 7602 Matieland, Stellenbosch, South Africa}
\affiliation{iThemba Laboratory for Accelerator Based Sciences, Somerset West 7129, South Africa}

\author{D. J. Mar\'{i}n-L\'{a}mbarri}
 \affiliation{iThemba Laboratory for Accelerator Based Sciences, Somerset West 7129, South Africa}
\affiliation{Department of Physics, University of the Western Cape, Private Bag X17, Bellville 7535, South Africa}
\affiliation{Instituto de F\'{i}sica, Universidad Nacional Aut\'{o}noma de M\'{e}xico, Apartado Postal 20-364, 01000 Cd. M\'{e}xico, M\'{e}xico}

\author{N. J. Mukwevho}
\affiliation{Department of Physics, University of the Western Cape, Private Bag X17, Bellville 7535, South Africa}

\author{F. Nemulodi}
\affiliation{Department of Physics, Stellenbosch University,
Private Bag X1, 7602 Matieland, Stellenbosch, South Africa}
\affiliation{iThemba Laboratory for Accelerator Based Sciences, Somerset West 7129, South Africa}

\author{R. Neveling}
\affiliation{iThemba Laboratory for Accelerator Based Sciences, Somerset West 7129, South Africa}

\author{P. Papka}
\affiliation{Department of Physics, Stellenbosch University,
Private Bag X1, 7602 Matieland, Stellenbosch, South Africa}
\affiliation{iThemba Laboratory for Accelerator Based Sciences, Somerset West 7129, South Africa}

\author{L. Pellegri}
\affiliation{iThemba Laboratory for Accelerator Based Sciences, Somerset West 7129, South Africa}
\affiliation{School of Physics, University of the Witwatersrand, Johannesburg 2050, South Africa}

\author{V. Pesudo}
 \affiliation{iThemba Laboratory for Accelerator Based Sciences, Somerset West 7129, South Africa}
\affiliation{Department of Physics, University of the Western Cape, Private Bag X17, Bellville 7535, South Africa}
 \affiliation{Centro de Investigaciones Energéticas, Medioambientales y Tecnológicas, Madrid 28040, Spain}

\author{B.M. Rebeiro}
\affiliation{Department of Physics and Astronomy, University of the Western Cape, Private Bag X17, Bellville 7535, South Africa}

\author{G. F. Steyn}
\affiliation{iThemba Laboratory for Accelerator Based Sciences, Somerset West 7129, South Africa}

\author{W. Yahia-Cherif}
\affiliation{University of Sciences and Technology Houari Boumedienne (USTHB),Faculty of Physics, P.O. Box 32, EL Alia, 16111 Bab Ezzouar, Algiers, Algeria}

\date{\today}

\begin{abstract}
\begin{description}
\item[Background] Type I X-ray bursts provide an opportunity to constrain the equation of state of nuclear matter. Observations of the lightcurves from these bursts allow the compactness of neutron stars to be constrained. However, the behaviour of these lightcurves also depends on a number of important thermonuclear reaction rates. One of these reactions, $^{18}$Ne($\alpha,p$)$^{21}$Na, has been extensively studied but there is some tension between the rate calculated from spectroscopic information of states above the $\alpha$-particle threshold in $^{22}$Mg and the rate determined from time-reversed measurements of the cross section.

\item[Purpose] The time-reversed measurement of the cross section is only sensitive to the ground state-to-ground state contribution. Therefore, corrections must be made to this reaction rate to account for the contribution of branches to excited states in $^{21}$Na. At present this is done with statistical models which may not be applicable in such light nuclei. Basing the correction of the time-reversed cross section on experimental data is much more robust.

\item[Method] The $^{24}$Mg($p,t$)$^{22}$Mg reaction was used to populate states in $^{22}$Mg. The reaction products from the reaction were analysed by the K600 magnetic spectrometer at iThemba LABS, South Africa. Protons decaying from excited states of $^{22}$Mg (S$_{p}$ = 5502 keV) were detected in an array of five double-sided silicon strip detectors placed at backward angles. The branching ratio for proton decays to the ground state of $^{21}$Na, $B_{p_0}$, was determined by comparing the inclusive (triton-only focal-plane) and exclusive (focal-plane gated on a specific proton decay) spectra.

\item[Results] The experimental proton decay branching ratio to the ground state of $^{21}$Na from excited states in $^{22}$Mg were found to be a factor of about two smaller than the ratios predicted by Hauser-Feshbach models. Using the experimental branchings for a recalculation of the $^{18}$Ne($\alpha$,p$_0$)$^{21}$Na cross section leads to a considerably improved agreement with previous reaction data. Updated information on the disputed number of levels around $E_x\approx9$ MeV
and on the possible $^{18}$Ne($\alpha$,2p)$^{20}$Ne cross section at astrophysical energies is also reported. 

\item[Conclusions] The proton decay branching of excited states in $^{22}$Mg to the ground state of $^{21}$Na have been measured using the K600 Q2D spectrometer at iThemba LABS coupled to the double-sided silicon-strip detector array CAKE. Using these experimental data, the modeling of the $^{18}$Ne($\alpha$,p$_0$)$^{21}$Na cross section has been improved. The result is not only in better agreement with previous cross section data but also consistent with a recent direct measurement of $^{18}$Ne($\alpha$,p)$^{21}$Na. This strengthens the case for the application of statistical models for these reactions.
\end{description}
\end{abstract}

\maketitle


\section{\label{sec:astrobg}Astrophysical Background}

Type I X-ray bursts are thermonuclear explosions which take place on the surface of neutron stars in binary systems \cite{grindlay1976discovery,woosley1976gamma}. Hydrogen- and helium-rich material from the companion star accretes onto the surface of the neutron star and hydrogen burning through the cold and hot CNO cycles commences. As the temperature keeps rising, eventually breakout from the CNO cycles can begin, increasing the rate of energy generation and resulting in the burst. The nucleosynthesis of the burst proceeds through a series of $\alpha$ particle- and proton-induced reactions, terminating in the SnSbTe cycle above $^{100}$Sn \cite{schatz2001end}.

The properties of neutron stars are the focus of a great deal of attention with the recent observation of neutron-star mergers through gravitational waves \cite{abbott2018gw170817} and the subsequent observation of the electromagnetic counterpart \cite{Abbott_2017}. One as-yet unanswered question is the equation of state of nuclear matter, and the subsequent relationship between the mass and radius of neutron stars (see e.g. Refs. \cite{lattimer2012nuclear}). Simulated X-ray burst lightcurves have shown a dependence on the compactness of neutron stars \cite{PhysRevLett.125.202701}. Constraints on the compactness from  the lightcurve can subsequently be used to limit the range of symmetry-energy parameters of nuclear matter. However, the lightcurve is driven by nuclear processes and unless the rates of the reactions driving the burst are known it is not possible to provide useful constraints on the neutron-star compactness.

Sensitivity studies by Meisel {\it et al.} \cite{Meisel_2019} and Cyburt {\it et al.} \cite{Cyburt_2016} have identified a number of important reactions which influence the shape of the X-ray burst lightcurve. The reactions identified (listed here in order of importance) were \cite{Meisel_2019}: $^{15}$O($\alpha,\gamma$)$^{19}$Ne, $^{14}$O($\alpha,p$)$^{17}$F, $^{23}$Al($p,\gamma$)$^{24}$Si, $^{59}$Cu($p,\gamma$)$^{60}$Zn, $^{18}$Ne($\alpha,p$)$^{21}$Na, $^{24}$Mg($\alpha,\gamma$)$^{28}$Si, $^{22}$Mg($\alpha,p$)$^{25}$Al, and $^{61}$Ga($p,\gamma$)$^{62}$Se. Recent developments have constrained some of these reaction rates \cite{PhysRevLett.122.232701,PhysRevLett.125.202701,PhysRevC.102.015801,PhysRevC.92.035801,PhysRevC.86.025801,PhysRevLett.98.242503,PhysRevC.72.041302,PhysRevC.102.045802,PhysRevC.77.055801}. For example, the $^{22}$Mg($\alpha,p$)$^{25}$Al cross section has been measured using a time-projection chamber at NSCL \cite{PhysRevLett.125.202701}. The results from this experiment suggest that the $^{22}$Mg($\alpha,p$)$^{25}$Al reaction rate is around 8 times smaller than the prediction from the statistical model NON-SMOKER \cite{NONSMOKER_web} and considerably higher than the rate calculated by Matic {\it et al.} \cite{PhysRevC.84.025801}, though the Matic rate was based on only four resonances and should correctly be regarded as a lower limit.

The $^{18}$Ne($\alpha,p$)$^{21}$Na reaction, which influences the shape of the rise of the X-ray burst lightcurve \cite{Meisel_2019}, has been the focus of extensive study. This includes time-reversed measurements of the $^{21}$Na($p,\alpha_0$)$^{18}$Ne cross section \cite{PhysRevLett.108.242701} (which will be discussed in more detail in the following paragraphs), direct measurements of the cross section using beams of radioactive $^{18}$Ne ions and gaseous $^{4}$He targets \cite{PhysRevC.66.055802,PhysRevC.59.3402,ANASEN}, transfer spectroscopy of $^{22}$Mg states using the $^{24}$Mg($p,t$)$^{22}$Mg reaction \cite{PhysRevC.80.055804,PhysRevC.79.055804}, and resonance-scattering $^{21}$Na($p,p$)$^{21}$Na measurements in inverse kinematics \cite{PhysRevC.88.012801,PhysRevC.89.015804}.

Of particular note for the currently reported measurement is the use of time-reversed measurements. Due to the technical difficulty of measuring the $^{18}$Ne($\alpha,p$)$^{21}$Na direct reaction requiring both an intense radioactive $^{18}$Ne beam and a helium gas target, Salter {\it et al.} \cite{PhysRevLett.108.242701} measured the time-reversed $^{21}$Na($p,\alpha$)$^{18}$Ne cross section. This has the advantage of allowing a solid hydrogen-containing target (e.g. CH$_2$) to be used in the experiment. However, this reaction is sensitive only to the $^{18}$Ne($\alpha,p_0$)$^{21}$Na cross section, i.e. reactions proceeding to the ground state of $^{21}$Na. In the Gamow window of the $^{18}$Ne($\alpha,p$)$^{21}$Na reaction at temperatures relevant to X-ray bursts, a large number of states in $^{21}$Na may be populated. Salter {\it et al.} suggest that this missing flux would not increase the cross section by more than a factor of three, as determined by the proton branching ratio of $^{22}$Mg to the ground state of $^{21}$Na calculated with the statistical model NON-SMOKER of Rauscher \cite{RAUSCHER200147,NONSMOKER_web}. 


Mohr and Matic \cite{PhysRevC.87.035801} compiled a detailed and valuable summary of existing data related to the $^{18}$Ne($\alpha,p$)$^{21}$Na reaction. 
They make detailed comparisons between three sets of data: the direct measurements of Groombridge {\it et al.} and Bradfield-Smith {\it et al.} \cite{PhysRevC.66.055802,PhysRevC.59.3402}, the time-reversed measurement of Salter {\it et al.} \cite{PhysRevLett.108.242701} and Sinha {\it et al.}, a time-reversed measurement only available in an internal report from Argonne National Laboratory \cite{SinhaANLReport}, and the transfer measurements of Matic {\it et al.} \cite{PhysRevC.80.055804}, and Chae {\it et al.} \cite{PhysRevC.79.055804}. They conclude that the cross sections from the direct measurements are problematically high, exhausting or exceeding the theoretical strengths derived by considering the states observed in the transfer measurements, and the cross section from the time-reversed experiment. This is true despite the widths from the Matic \cite{PhysRevC.80.055804} $^{24}$Mg($p,t$)$^{22}$Mg data being used to compute resonance strengths in contrast to the larger widths from the direct measurement of the $^{18}$Ne($\alpha,p$)$^{21}$Na reaction rate.

The recommended reaction rate of Mohr and Matic \cite{PhysRevC.87.035801} exceeds that determined from the time-reversed measurement of Salter {\it et al.} by around a factor of 5. However, this comparison is for the ground state-to-ground state reaction rate. After taking into account the contribution of transitions to excited states, the Salter {\it et al.} reaction rate is a factor of 2-3 times smaller than the recommended reaction rate. Mohr and Matic state that the conclusion of Salter {\it et al.} that ``the breakout from the Hot CNO cycles via the $^{18}$Ne($\alpha,p$)$^{21}$Na reaction is delayed and occurs at higher temperatures than previously predicted'' is not supported in the current calculations. The comparison with the Salter {\it et al.} data was performed by using a uniform factor of 3 to account for the decay branching of the $^{22}$Mg excited states to the ground state of $^{21}$Na.

A subsequent study by Mohr, Longland and Iliadis \cite{PhysRevC.90.065806} provided a new evaluated rate based on the known states in $^{22}$Mg. This rate used the {\sc RatesMC} Monte Carlo code \cite{ILIADIS2010251,Sallaska_2013} to estimate the reaction rate with realistic uncertainties. The rate which resulted from these calculations was somewhat reduced compared to the earlier evaluation of Mohr and Matic \cite{PhysRevC.87.035801}, but in better agreement with the rate derived from the time-reversed experiment of Salter {\it et al.} \cite{PhysRevLett.108.242701}. The calculation of the reaction rate also resulted in a relatively well-constrained temperature ($0.60(2)$ GK) at which the $^{18}$Ne($\alpha,p$)$^{21}$Na reaction becomes faster than the $\beta^+$ decay of $^{18}$Ne.

In this paper we report on a study of the $^{24}$Mg($p,t$)$^{22}$Mg($p$)$^{21}$Na reaction in an attempt to resolve any discrepancies between the rate calculated from spectroscopic information of states above the $\alpha$-particle threshold in $^{22}$Mg and the rate determined from time-reversed measurements of the cross section. This reaction was measured in an experiment with the K600 magnetic spectrometer at iThemba LABS, South Africa coupled to the \textsc{cake} (Coincidence Array for K600 Experiments), an array of five double-sided silicon strip detectors. The proton decays from the populated excited $^{22}$Mg states were observed and, with this information, the contribution of proton decays to excited states in $^{21}$Na to the $^{18}$Ne($\alpha,p$)$^{21}$Na reaction rate was reassessed. In light of the obtained energy-dependent branching ratio, we re-estimate the $^{18}$Ne($\alpha,p$)$^{21}$Na reaction rate based on the time-reversed measurement of Salter {\it et al.} \cite{PhysRevLett.108.242701} and compare it to other evaluations of the reaction rate based on spectroscopic information on $^{22}$Mg.

\section{Experimental Method}

A dispersion-matched beam of 100-MeV protons was extracted from the Separated-Sector Cyclotron at iThemba LABS and transported along a dispersion-matched beamline to the target position of the K600 magnetic spectrometer, a kinematically corrected Q2D spectrometer \cite{NEVELING201129}. For the measurement of excited $^{22}$Mg states, one $^{24}$Mg target was used of $750$ $\mu$g/cm$^2$. Background data were taken with $^{12}$C and mylar targets. The K600 spectrometer was placed so that the acceptance aperture was located at zero degrees. Unreacted beam passed into the K600 spectrometer and was stopped on a Faraday cup located within the first dipole.

\begin{figure}
\includegraphics[width=0.48\textwidth]{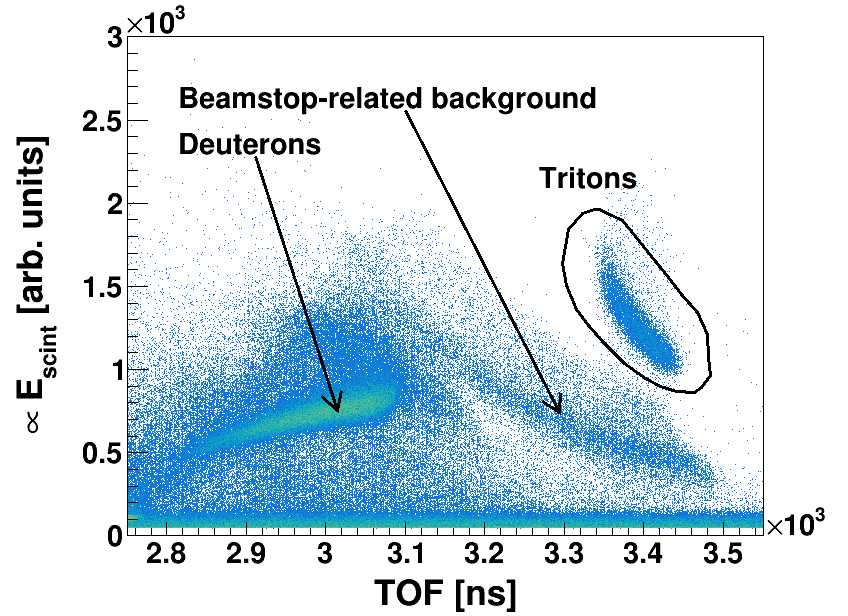}
 \caption{Particle identification for the tritons in the K600 magnetic spectrometer. The triton locus is encircled by a black border which is representative of the software gate used. Most of the other events are caused by beamstop-induced background or by deuterons caused by reactions in the target.}
 \label{fig:PID}
\end{figure}

Reaction products were momentum-analysed in the K600 magnetic spectrometer and detected in a focal-plane detector suite consisting of two drift chambers and two plastic scintillators placed at the medium-dispersion focal plane. The trigger for the experiment was an event in the first plastic scintillator. Tritons were identified by considering the time-of-flight through the K600 relative to an RF reference value and the energy deposition in the plastic scintillator (see Fig. \ref{fig:PID}).

Protons decaying from $^{22}$Mg excited states were detected in the Coincidence Array for K600 Experiments (the \textsc{cake}), an array of five double-sided silicon strip detectors placed at backward angles within the scattering chamber \cite{adsley2017cake}. The signals from the silicon detectors were fed into amplifier modules which provided shaped Gaussian signal outputs which were processed by amplitude-to-digital convertor (ADC) modules in the DAQ. The time signal was generated from a constant-fraction discriminator which was acquired by time-to-digital (TDC) convertor modules in the DAQ. The silicon time values were all recorded relative to the trigger from the K600 focal plane which is used as a reference time in all TDC modules to obviate the impact of any clock jitter between different TDC modules.

\section{Data Analysis}

The data were processed, constructing all events relating to the focal plane e.g. generating the position spectra from the drift chamber timing values. The condition of a good event from the \textsc{cake} was when timing value in a front silicon strip was found along with a recorded energy in the same front strip for which an approximately equal recorded energy ($|E_\mathrm{front} - E_\mathrm{back}| < 0.2$ MeV) in a back silicon strip in the same detector was found. The data analysis was truncated for $E{_x} > 12$ MeV.

\begin{figure}
\includegraphics[width=0.48\textwidth]{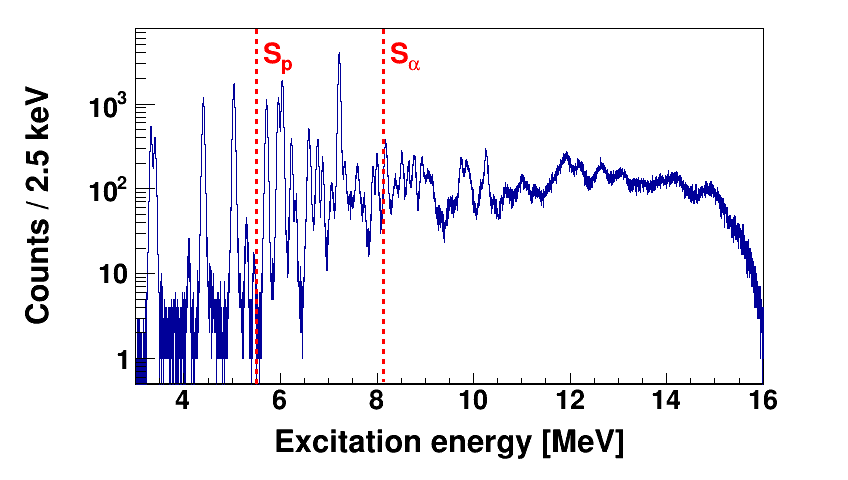}
 \caption{Inclusive excitation-energy spectrum of levels in $^{22}$Mg. The proton-decay threshold (S$_{p}$ = 5.502 MeV) and the $\alpha$-particle threshold (S$_{\alpha}$ = 8.142 MeV) is indicated.}
 \label{fig:ExSpectrumInclusive}
\end{figure}

The resulting inclusive (triton-gated focal-plane only, ``singles") excitation-energy spectrum for the combined dataset is shown in Fig. \ref{fig:ExSpectrumInclusive}. The excitation-energy resolution is $45(2)$ keV (FWHM), this is significantly larger than the $15$ keV (FWHM) \cite{PhysRevC.80.055804} achieved at RCNP Osaka with the Grand Raiden, due to the dispersion matching of the beam and the K600 fields not remaining stable during the experiment. The excitation energy is always determined from the focal plane and so the energy resolution is the same for inclusive and exclusive spectra.

\begin{figure}
\includegraphics[width=0.48\textwidth]{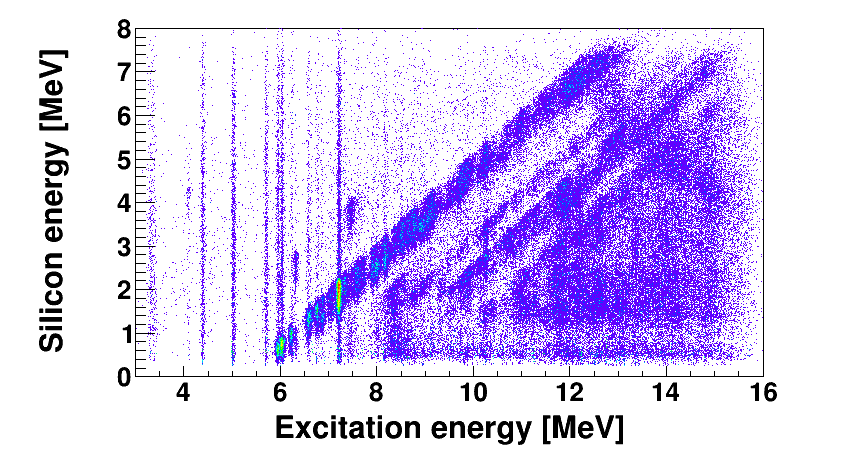}
\includegraphics[width=0.48\textwidth]{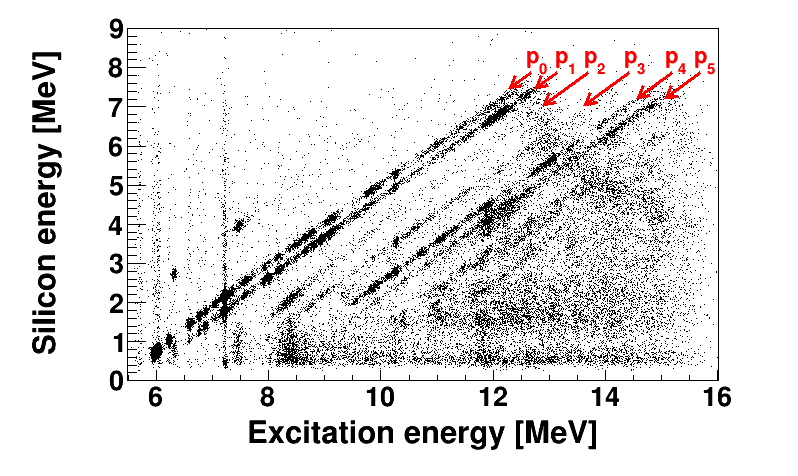}
 \caption{Excitation energy vs. silicon energy coincidence spectrum. (Top) Combined coincidence spectrum (for reference), the broadening of the coincidence loci is due to the effective target thickness change experienced by the decaying proton. (Bottom) Coincidence spectrum selected on a small number of angular rings (ring 1 ($\theta_{lab} = 161.4^{\text{o}}$) to ring 4 ($\theta_{lab} = 151.4^{\text{o}}$)) on the silicon detectors; the coincidence loci are much narrower due to the smaller range of effective target thicknesses. Each decay locus from $p_0$ to $p_5$ is indicated. The contaminants seen in the kinematically-inaccessible region, to the left of the $p_0$ decay locus, are from oxygen.}
 \label{fig:ExEsi2D}
\end{figure}

Two excitation energy vs. silicon energy coincidence spectra are shown in Fig. \ref{fig:ExEsi2D}. The top panel shows the combined coincidence spectrum; the spread in the coincidence loci in this two-dimensional spectrum is predominantly due to the different effective target thickness for the decaying proton. The effective thickness of the target as viewed by the different rings of the \textsc{cake} increases from 779 $\mu$g/cm$^2$ to 1839 $\mu$g/cm$^2$ as the angles change from 114$^{\text{o}}$ to 164$^{\text{o}}$. The combined spectrum results from chaining all the data runs together and plotting the combined data on one spectrum. The vertical lines are caused by weak, random coincidences with the strong resonances in $^{22}$Mg. The bottom panel shows a coincidence spectrum for a limited number of angular rings in the silicon detectors. A selection of decay channels are done on subsets of rings to give clear separation of these decay channels.

Coincidence spectra gated on the $p_0$, $p_1$, $p_2$, $p_3$, $p_4$, and $p_5$ ($p_i$ meaning the proton decay to the $i$th excited state in $^{21}$Na) decay loci were generated by gating on the decay loci in these two-dimensional spectra. These correspond to proton decays to the ground state and various excited states in $^{21}$Na from the recoil nucleus $^{22}$Mg and are listed in Table \ref{table:21na-levels}. Higher decay loci were not analysed as the amount of background relative to the signal for these decay channels was too high to place software gates around them. The spectra were limited to $E_\mathrm{x} < 12$ MeV; above this energy the protons decaying to the ground state of $^{21}$Na punched through the 400-$\mu$m-thick silicon detectors comprising the \textsc{cake}.

\begin{table}[htbp]
\centering
\caption{Levels in $^{21}$Na (S$_{p}$ = 2432 keV) that are populated due to proton decay from $^{22}$Mg. The $p_{4}$ and $p_{5}$ decay modes proceed to proton-unbound states which decay by $\gamma$-ray emission (see the text). The excitation energies (and uncertainties) and spin-parities are taken from Ref. \cite{ENSDF}.}
\begin{ruledtabular}
\begin{tabular}{ c c c c} 
 Decay & E$_{x}$ [keV] & J$^{\pi}$ & $E_\mathrm{x} - S_{p,^{21}\mathrm{Na}}$ [keV] \\
 \hline
 $p_{0}$ & (g.s.) & $3/2^+$ & (bound) \\ 
 $p_{1}$ & $331.90(10)$ & $5/2^+$ & (bound) \\ 
 $p_{2}$ & $1716.1(3)$ & $7/2^+$ & (bound) \\ 
 $p_{3}$ & $2423.8(4)$ & $1/2^+$ & (bound) \\
 $p_{4}$ & $2797.9(5)$ & $1/2^-$ & $366.22(57)$ \\
 $p_{5}$ & $2829.1(7)$ & $9/2^+$ & $397.42(75)$ \\
\end{tabular}
 \end{ruledtabular}
\label{table:21na-levels}
\end{table}

\begin{figure*}
\includegraphics[width=0.95\textwidth]{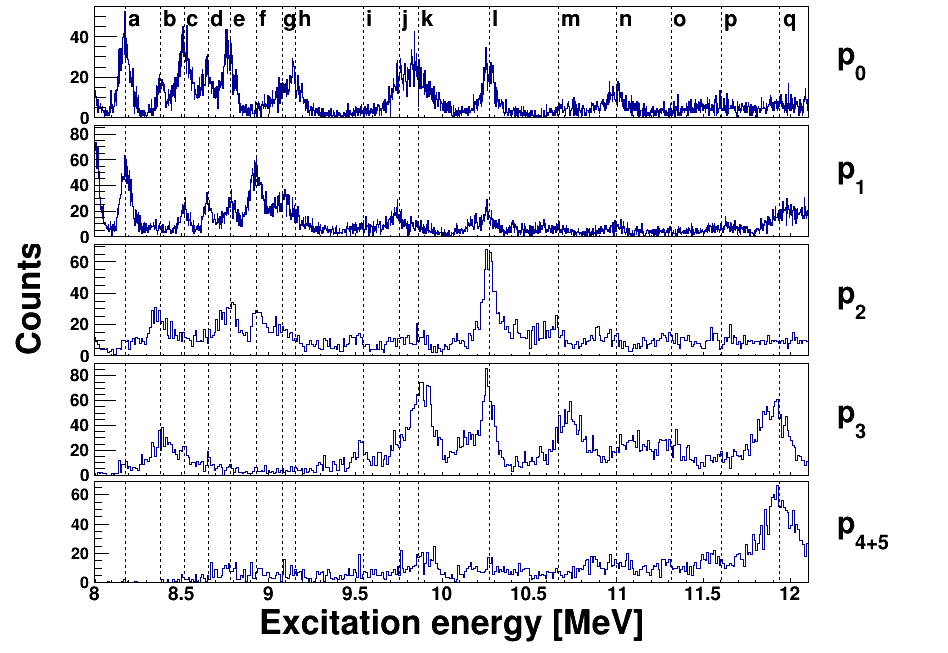}
 \caption{Exclusive (gated on specific proton decays to different final $^{21}$Na states) excitation-energy spectra. These spectra are gated on the $p_0$ (top) to $p_{4+5}$ (bottom) decay channels. The $p_{4+5}$ spectrum includes the $p_4$ and $p_5$ decay locus as they are 30 keV apart and not resolved. The $p_0$ and $p_1$ spectra have a bin size of 2.5 keV, the other spectra have 10-keV binning. The states that correspond to RCNP data from Matic {\it et al.} \cite{PhysRevC.84.025801} are labeled in the top panel. Table \ref{table:22Mg-levels} lists the details of each state that corresponds to each label in the top panel of this figure.}
 \label{fig:ExExclusive}
\end{figure*}

The generated coincidence spectra, shown in the panels of Fig. \ref{fig:ExExclusive}, were analysed together with the inclusive excitation-energy spectrum given in Fig. \ref{fig:ExSpectrumInclusive}. Below the proton-decay threshold, where a number of isolated states could be observed, the analysis could proceed on a state-by-state basis. Above the proton threshold this was not possible due to the limited excitation-energy resolution and the low cross section of reactions to the excited states. Instead, we adopted a similar approach to that of Munson {\it et al.} \cite{munson} who performed analysis of excitation-energy bins rather than individual states. In that experiment, states in $^{24}$Mg were populated through the inelastic scattering of $\alpha$ particles from a $^{24}$Mg target. With a relatively coarse binning, the decay of each excitation-energy bin into the available final channels was considered rather than a state-by-state analysis. The aim is to characterise the overall trend of the $p_0$ and other branching ratios in order to evaluate a realistic range for the $B_{p_0}$ as a function of excitation energy. The size of the excitation-energy bins used for this analysis technique was 100 keV.

\begin{table}[htbp]
\centering
\caption{Excitation energies for states above $S_\alpha$ corresponding to the labels as seen in Fig. \ref{fig:ExExclusive}. These are from Matic {\it et al.} \cite{PhysRevC.80.055804} and agree well with the states identified in this study. The updated spin-parity assignments are from Table I in Mohr, Longland, and Iliadis \cite{PhysRevC.90.065806}.}
\begin{ruledtabular}
\begin{tabular}{ c c c c } 
 Label & E$_{x}$ [MeV] & Matic E$_{x}$ [MeV] & J$^{\pi}$ \\
 \hline
 a & 8.1721(20) & 8.1803(17)  & $2^+$\\
 b & 8.3886(15) & 8.383(13)   & $1^+$\\
 c & 8.520(13) & 8.5193(21)  & $3^-$\\
 d & 8.652(9) & 8.6575(17)  & $2^+$\\
 e & 8.777(22) & 8.743(14)   & $1^-$\\
 f & 8.937(19) & 8.9331(29)  & $2^+$\\
 g & 9.067(23) & 9.082(7)    & $1^-$\\
 h & 9.147(15) & 9.157(4)    & $4^+$\\
 i & 9.537(11) & 9.546(15)   & $1^-$\\
 j & 9.745(16) & 9.7516(27)  & $2^+$\\
 k & 9.873(19) & 9.861(6)    & $0^+$\\
 l & 10.259(20) & 10.2717(17) & $2^+$\\
 m & 10.660(8) & 10.667(19)  & $3^-$\\
 n & 10.998(6) & 10.999(15)  & $4^+$\\
 o & 11.309(14) & 11.317(27)  & $4^+$\\
 p & 11.547(29) & 11.603(16)  & $4^+$\\
 q & 11.931(13) & 11.937(17)  & $0^+$\\
\end{tabular}
 \end{ruledtabular}
\label{table:22Mg-levels}
\end{table}

\begin{figure*}
\includegraphics[width=0.95\textwidth]{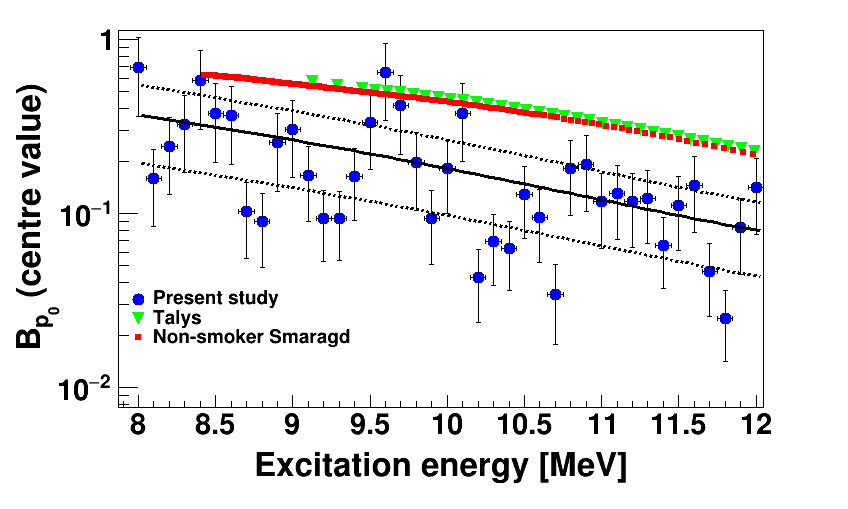}
\includegraphics[width=1.0\textwidth]{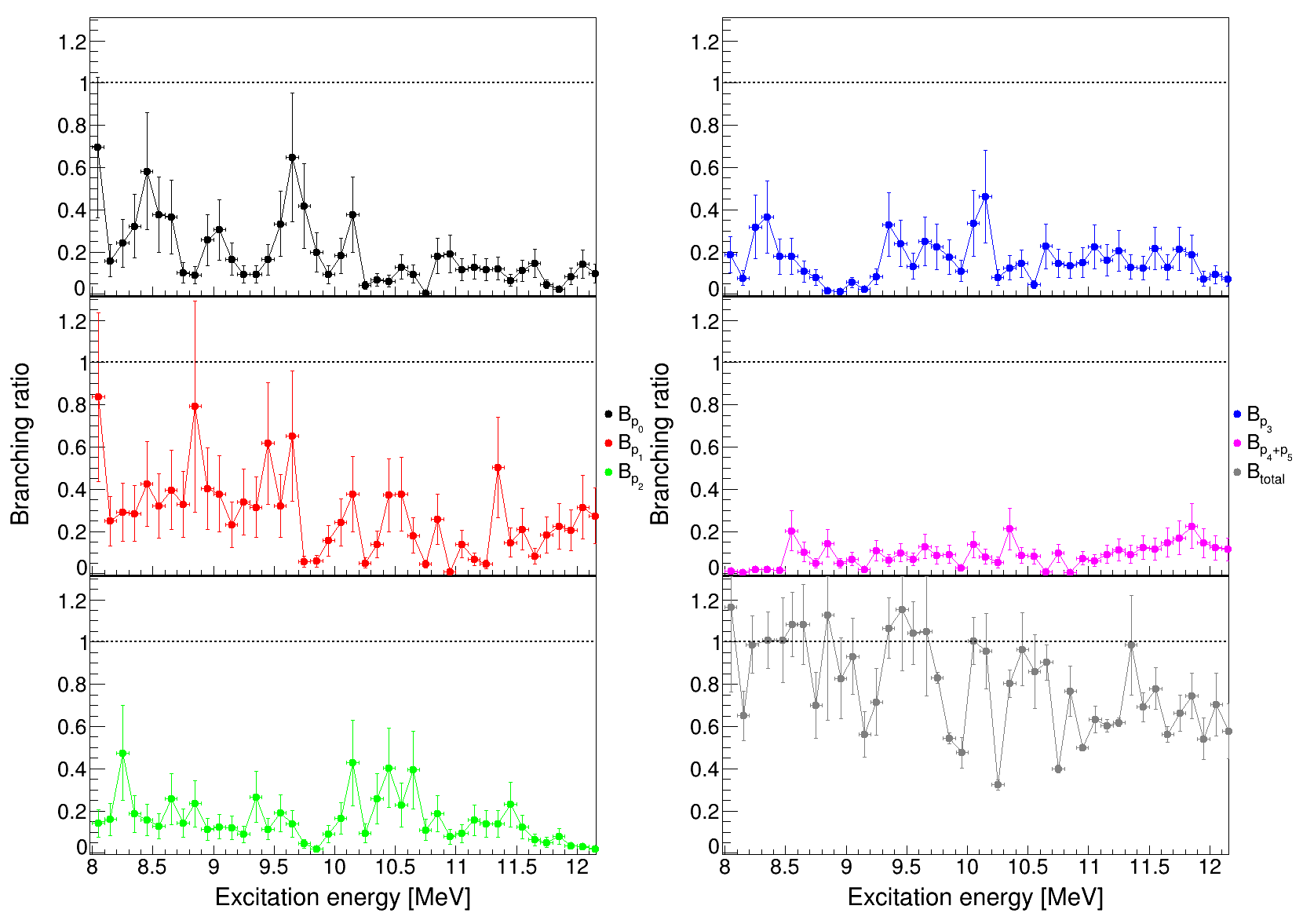}
 \caption{(color online) (Top) Centre-value $B_{p_{0}}$ as a function of $E_\mathrm{x}$ (blue) with quadratic fit to the data (black). The uncertainties are not indicated on individual $B_{p_{0}}$ data points in the upper panel but are present in the lower panel. Quadratic fits to the upper and lower limits of the $B_{p_{0}}$ branching ratios are shown as black-dotted lines. $B_{p_{0}}$ for the $^{18}$Ne($\alpha,p$)$^{21}$Na reaction calculated with TALYS is shown in green. The NON-SMOKER $B_{p_{0}}$ using the SMARAGD model is shown by the red line. (Bottom) $B_{p_i}$ and the sum of the proton branching ratios as a function as excitation energy. The bin size is 100 keV.} 
 \label{fig:BranchingRatios}
\end{figure*}

The coincidence yields for each excitation-energy bin were corrected for the missing solid angle by fitting the yields per ring with a linear combination of the even Legendre polynomials. The $p_0$ branching ratio of each of the excitation-energy bins is shown in the top panel of Fig. \ref{fig:BranchingRatios}. The bottom panel of Fig. \ref{fig:BranchingRatios} shows the sum of the branching ratios for $p_0$ to $p_5$, inclusive. This shows that the total branching ratio, especially at lower excitation energies sums to 1. While there is significant variation of the branching ratios, especially at lower excitation energies, the general trend of $B_{p_0} \sim 0.4$ at $E_\mathrm{x} \sim 8$ MeV decreasing to $B_{p_0} \sim 0.1$ at higher excitation energies ($E_\mathrm{x} \sim 12$ MeV) is clear. However, there is a significant range of $B_{p_0}$ values observed in the present experiment. In Section \ref{sec:discussion} we set out the implications of the newly determined branching ratios for the validity of statistical models in this system, and on the reaction rate determined from the time-reversed measurements of Salter {\it et al.} \cite{PhysRevLett.108.242701}. Table \ref{tab:DecaysToFinalNa21Levels} gives a summary of the observed decays from $^{22}$Mg excited states to the various final states in $^{21}$Na.

\begin{table}[htbp]
    \centering
    \caption{Observed decays of $^{22}$Mg states into $^{21}$Na states. The first column shows the label for the $^{22}$Mg state which corresponds to the energies listed in Table \ref{table:22Mg-levels}. The second column lists the $p_i$ decays observed for that particular state.}
    \begin{ruledtabular}
    \begin{tabular}{ c m{3cm} c m{6cm} }
        Label & Decay type \\
        \hline
        a & $p_0$, $p_1$  \\
        b & $p_0$, $p_2$, $p_3$ \\
        c & $p_0$, $p_1$, $p_2$, $p_3$ \\
        d & $p_0$, $p_1$, $p_2$, $p_3$ \\
        e & $p_0$, $p_1$, $p_2$, $p_{4,5}$  \\
        f & $p_1$, $p_2$ \\
        g & $p_0$, $p_1$, $p_2$ \\
        h & $p_0$, $p_1$ \\
        i & $p_2$, $p_3$ \\
        j & $p_0$, $p_1$, $p_3$ \\
        k & $p_0$, $p_3$, $p_{4,5}$ \\
        l & $p_0$, $p_1$, $p_2$, $p_3$ \\
        m & $p_2$, $p_3$ \\
        n & $p_0$, $p_3$, $p_{4,5}$ \\
        o & $p_2$, $p_3$, $p_{4,5}$ \\
        p & $p_{4,5}$ \\
        q & $p_1$, $p_3$, $p_{4,5}$ \\
    \end{tabular}
    \end{ruledtabular}
    \label{tab:DecaysToFinalNa21Levels}
\end{table}

The branching ratio determined is dependent on both the number of events detected in the $p_0$ coincidence spectrum but also on the number of events detected in the inclusive spectrum. Target contaminants, notably $^{12}$C and $^{16}$O lie on the focal plane close to the region of interest and could potentially influence the extracted branching ratios. Data were taken with $^{12}$C and Mylar targets to locate the contaminant states and quantify their potential impact on the branching ratios. The population of these states was observed to be rather weak compared to the $^{22}$Mg states, likely due to the rather thick $^{24}$Mg targets used in the experiment, and the impact of the contamination on the branching ratios is marginal ($<$1\%). 

\section{Discussion\label{sec:discussion}}

In this section, we first discuss the implications of the presently determined trends in the $p_0$ branching ratio of $^{22}$Mg states on the $^{18}$Ne($\alpha,p$)$^{21}$Na reaction rate based on the time-reversed measurements of the $^{21}$Na($p,\alpha$)$^{18}$Ne reaction. After this we make some brief comments about the number and nature of the levels around $E_\mathrm{x} \sim 9$ MeV based on the current data, since the number and nature of states at this excitation energy could strongly influence the $^{18}$Ne($\alpha,p$)$^{21}$Na reaction rate. Finally, we make a brief comment about the proton decay branches from $^{22}$Mg states to proton-unbound $^{21}$Na states which may result in the $^{18}$Ne($\alpha,2p$)$^{20}$Ne reaction contributing at astrophysical energies.

\subsection{Impact of p$_0$ branching ratios on the calculation of the $^{18}$Ne($\alpha$,p$_0$)$^{21}$Na cross 
section}
\label{sec:impact}

Salter \textit{et al.} \cite{PhysRevLett.108.242701} determined the reaction rate of $^{18}$Ne($\alpha$,p$_0$)$^{21}$Na from a measurement of the
$^{21}$Na(p,$\alpha$)$^{18}$Ne cross section and applying detailed balance. This determination, however, may not fully constrain the 
actual \textit{stellar} reaction rate for $^{18}$Ne($\alpha$,p)$^{21}$Na to be used in astrophysical simulations. To this end, 
particle transitions to and from excited states of the target and residual nuclei have to be considered additionally 
\cite{2020entn.book.....R}. The contribution of excited states in $^{18}$Ne to the stellar rate is negligible for temperatures 
attained in X-ray bursts ($\leq 3$ GK) \cite{2012ApJS..201...26R}. Transitions to excited states in $^{21}$Na, however, may affect 
the rate. These could not be constrained by the previous measurement \cite{PhysRevLett.108.242701} and so far could only be predicted by theory.

The proton-branching ratios obtained in this work are related to the ones found in the reaction $^{18}$Ne($\alpha$,p)$^{21}$Na only 
if the proton 
emission from the states in $^{22}$Mg proceeds in a similar manner in  $^{18}$Ne($\alpha$,p)$^{21}$Na and in our reaction sequence
$^{24}$Mg(p,t)$^{22}$Mg(p)$^{21}$Na. The $^{18}$Ne($\alpha$,p)$^{21}$Na and $^{24}$Mg(p,t)$^{22}$Mg are expected to produce rather 
different spin distributions in $^{22}$Mg and this could have some impact on the $B_\mathrm{p0}$ observed. However, most of the 
excitation energy bins considered in the present measurement are dominated by a single state. We assume that the Bohr hypothesis 
holds and the decay of a compound-nuclear state is insensitive to the formation channel. This implies that the decay of a $^{22}$Mg 
state
populated in either reaction will proceed identically. The low number of levels per bin and the insensitivity of the decay to the 
populating reaction further means that any spin-distribution effect has limited impact and may be ignored for further analysis.

The NON-SMOKER Hauser-Feshbach code reproduced the ($\alpha$,p$_0$) results of Salter \textit{et al.} acceptably well (see Fig.\ 2 
in \cite{PhysRevLett.108.242701}). 
Therefore we assume that the cross section can be described by a statistical model, despite the light target nucleus. The 
statistical model cross section $\sigma_{(\alpha,\mathrm{p})}$ is proportional to the averaged $\alpha$-width $\langle 
\Gamma_\alpha\rangle$ and the averaged proton-width $\langle \Gamma_\mathrm{p}\rangle$, and inversely proportional to the total 
width 
$\langle\Gamma_\mathrm{tot}\rangle$, \cite{2020entn.book.....R}
\begin{equation}
\label{eq:hfcs}
\sigma_{(\alpha,\mathrm{p})}\propto \frac{\langle\Gamma_\alpha\rangle\,\langle 
\Gamma_\mathrm{p}\rangle}{\langle\Gamma_\mathrm{tot}\rangle}\quad,
\end{equation}
with the particle widths being the sum of partial widths for transitions to the ground state and excited states: $\langle 
\Gamma_\alpha\rangle=\langle 
\Gamma_{\alpha 0}\rangle+\langle\Gamma_{\alpha 1}\rangle+\dots$ and $\langle 
\Gamma_\mathrm{p}\rangle=\langle\Gamma_\mathrm{p0}\rangle+\langle\Gamma_\mathrm{p1}\rangle+\dots$
The total width $\langle\Gamma_\mathrm{tot}\rangle=\langle 
\Gamma_{\alpha}\rangle+\langle\Gamma_\mathrm{p}\rangle+\langle\Gamma_\gamma\rangle$ further includes the
averaged $\gamma$-width $\langle\Gamma_\gamma\rangle$. For the case considered here, however, 
$\langle\Gamma_\gamma\rangle\ll\langle 
\Gamma_{\alpha}\rangle\ll\langle\Gamma_\mathrm{p}\rangle$ and can be neglected. Furthermore, 
$\langle\Gamma_\alpha\rangle=\langle\Gamma_{\alpha_0}\rangle$ for laboratory cross sections and also in stellar plasmas at 
temperatures below 3 GK because thermal population of excited states in $^{18}$Ne does not contribute to the stellar reaction rate.
The 
energy dependence of $\sigma_{(\alpha,\mathrm{p})}$ is mainly 
given by $\langle 
\Gamma_\alpha\rangle$ because $\langle 
\Gamma_\alpha\rangle\ll\langle\Gamma_\mathrm{p}\rangle$. Due to this relation and the fact that $\langle\Gamma_\mathrm{p}\rangle$ 
also enters the denominator in Eq.\ (\ref{eq:hfcs}), the cross section does not scale directly with a change in the proton width. 
Rather, the sensitivity of the cross section to a change in the proton width is small (for an in-depth discussion of such 
sensitivities, see \cite{2012ApJS..201...26R}). 

The measured branching ratio $B_\mathrm{p0}$ constrains the ratio 
$\langle\Gamma_\mathrm{p0}\rangle/\langle\Gamma_\mathrm{p}\rangle$. In order to use $B_\mathrm{p0}$ to improve the calculation of 
the ($\alpha$,p) reaction cross section, it is also necessary to either know $\langle\Gamma_\mathrm{p}\rangle$ or 
$\langle\Gamma_\mathrm{p0}\rangle$. This information is not available in the present data. Also the reproduction of the Salter 
\textit{et al.} data \cite{PhysRevLett.108.242701} for the ($\alpha$,p$_0$) reaction by the Hauser-Feshbach calculation is not suited to completely 
constrain these widths. This is because for the calculation of the ($\alpha$,p$_0$) cross section, the 
$\langle\Gamma_\mathrm{p}\rangle$ in the numerator of Eq.\ (\ref{eq:hfcs}) is replaced by $\langle\Gamma_\mathrm{p0}\rangle$ but 
in the denominator must still use a total width $\langle\Gamma_\mathrm{tot}\rangle$ calculated with the full proton width,
\begin{equation}
\label{eq:hfcs0}
\sigma_{(\alpha,\mathrm{p0})}\propto \frac{\langle\Gamma_\alpha\rangle\,
\langle\Gamma_\mathrm{p0}\rangle}{\langle\Gamma_\mathrm{tot}\rangle}\quad.
\end{equation}

The missing information on the averaged proton widths requires to apply further assumptions. The simplest assumption is that 
the total proton-width is predicted correctly and only the relation between $\langle\Gamma_\mathrm{p0}\rangle$ and 
$\langle\Gamma_\mathrm{p}\rangle$ is changed. This would affect the prediction of the ($\alpha$,p$_0$) cross section but not the 
one of the ($\alpha$,p) cross section.

For this calculation, we make use of the SMARAGD code 
\cite{2011IJMPE..20.1071R,SMARAGD}, which is the successor to the NON-SMOKER code. We verified beforehand that it produces identical results 
for this reaction with its default settings. As shown in Fig.\ 5, which compares the measured $B_\mathrm{p0}$ to 
$\sigma_{(\alpha,\mathrm{p0})}/\sigma_{(\alpha,\mathrm{p})}$ ratios from two different statistical model codes, there is good 
agreement between the theory results but about a factor of two difference to the average of the measured p$_0$ branchings, with 
theory predicting a larger contribution of $\langle\Gamma_\mathrm{p0}\rangle$ to the total proton width. There is, however, 
excellent agreement between experiment and theory in the energy dependence of the averaged branching.

\begin{figure}
\includegraphics[width=\columnwidth]{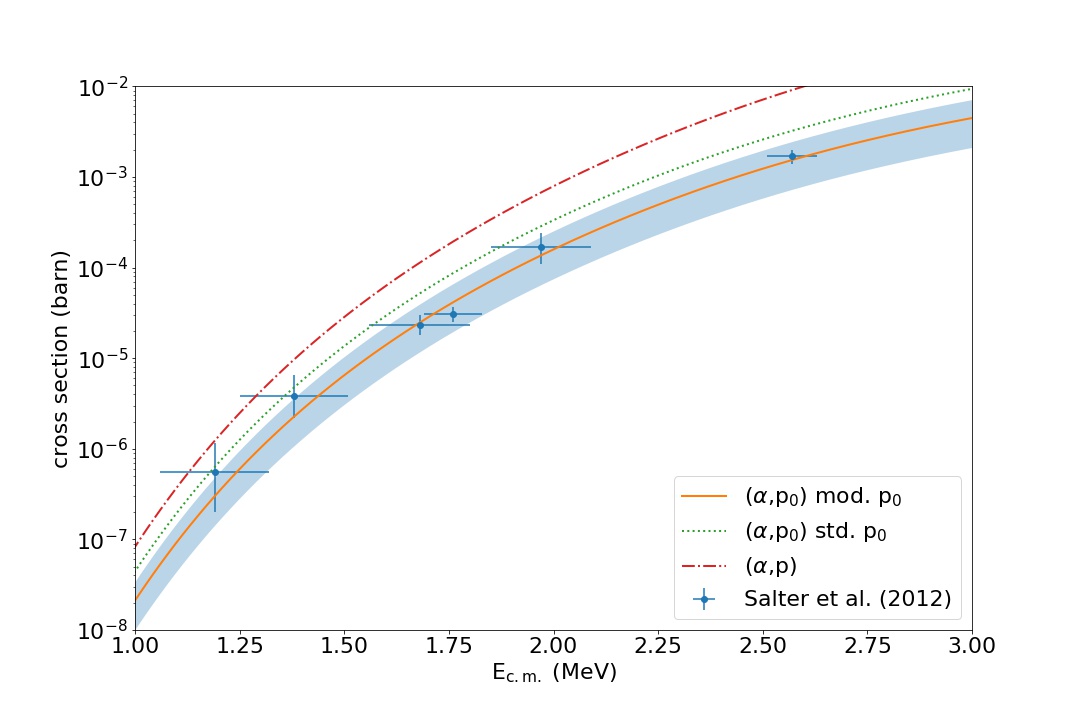}
\caption{\label{fig:hf_rescaled}Cross sections for $^{18}$Ne($\alpha$,p$_0$)$^{21}$Na and $^{18}$Ne($\alpha$,p$_0$)$^{21}$Na 
calculated with the SMARAGD code \cite{2011IJMPE..20.1071R,SMARAGD} with and without modified p$_0$ width. The shaded band shows 
the experimental uncertainty on the average p$_0$ branching.}
\end{figure}

Accounting for the 
uncertainty in the average branching as shown in Fig.\ 5, the predicted $\langle\Gamma_\mathrm{p0}\rangle$ has to be divided by 
$2.1_{-0.8}^{+2.4}$.
Figure \ref{fig:hf_rescaled} shows the results obtained when rescaling $\langle\Gamma_\mathrm{p0}\rangle$ but leaving 
$\langle\Gamma_\mathrm{p}\rangle$ unchanged. As expected from Eq.\ (\ref{eq:hfcs}), the ($\alpha$,p$_0$) cross section is reduced 
with respect to the calculation with the standard value $\langle\Gamma_\mathrm{p0}\rangle$. This leads to a considerably improved 
agreement with the ($\alpha$,p$_0$) data. The calculated ($\alpha$,p) cross section remains unchanged due to the adopted assumption.

\subsection{States at $E_\mathrm{x} \sim 9$ MeV}

The number and properties of the $^{22}$Mg levels at $E_\mathrm{x} \sim 9$ MeV are rather important since an additional state at this energy could potentially significantly increase the reaction rate, and therefore the ignition conditions for X-ray bursts. Previously there was some disagreement about the properties of levels here. The ENSDF databased \cite{ENSDF} gives $E_\mathrm{x} = 8891(7)$ keV for a state with a tentative $J^\pi = 1^-$ assignment. The $J^\pi = 1^-$ assignment comes from the $^{21}$Na($p,p$)$^{21}$Na resonance scattering experiments of He {\it et al.} and Zhang {\it et al.} \cite{PhysRevC.88.012801,PhysRevC.89.015804} in which a state is observed at $E_\mathrm{x} = 9.050(30)$ MeV. The listed energy is the weighted average of a $E_\mathrm{x} = 8985(8)$-keV level observed in proton decays of $^{22}$Mg populated in the $\beta$ decay of $^{22}$Al \cite{achouri2006beta} and a $E_\mathrm{x} = 9029(20)$-keV level observed in the $^{24}$Mg($\alpha,^6$He)$^{22}$Mg reaction \cite{BERG2003608,shimizu2005resonance}. This latter state, and the resonance observed in the $^{21}$Na($p,p$)$^{21}$Na scattering experiment, is presumably the $E_\mathrm{x} = 9080(7)$-keV level observed in the high-resolution $^{24}$Mg($p,t$)$^{22}$Mg data of Matic {\it et al.} \cite{PhysRevC.80.055804} since both of these reactions are two-neutron removal reactions and the difference between these energy levels is consistent with the observed scatter in energy levels between the two datasets.

The previous concern was whether the state at $E_\mathrm{x} = 9.318(12)$ MeV, observed in the $^{24}$Mg($p,t$)$^{22}$Mg reaction, undergoing a proton decay to the first-excited $332$-keV state in $^{21}$Na, could produce a proton peak at $E_p = 3.484(8)$ MeV which was misassigned in the decay study of Achouri {\it et al.} \cite{achouri2006beta} as a decay to the ground state of $^{21}$Na resulting in a spurious $E_\mathrm{x} = 8985(8)$-keV state in $^{22}$Mg. The $\beta$ decay study of Wu {\it et al.} \cite {PhysRevC.104.044311} has found a proton peak at $E_p = 3.511(11)$ MeV which corresponds to the $E_p = 3.484(8)$-MeV proton peak of Achouri {\it et al.} \cite{achouri2006beta}. However, in Ref. \cite {PhysRevC.104.044311} this proton peak was found to be in coincidence with a $E_\gamma = 332$-keV $\gamma$ ray. This suggests that, contrary to the existing assumptions on the levels in this region, there is only one at $E_\mathrm{x} = 9080(7)$ keV, and the previously assigned $E_\mathrm{x} = 8985(8)$-keV level should be removed and replaced by a level at $E_\mathrm{x} = 9317(8)$ keV. This satisfies both the most recent $^{22}$Al $\beta$-decay measurement \cite{PhysRevC.104.044311} and the high-resolution $^{24}$Mg($p,t$)$^{22}$Mg measurement of Ref. \cite{PhysRevC.80.055804}.

\subsection{Decays to proton-decay $^{21}$Na states}

The low proton threshold in $^{21}$Na leaves open the possibility that the $^{18}$Ne($\alpha,2p$)$^{20}$Ne reaction could be active within the Gamow window. We are able to test assumptions about this reaction channel by inspecting our data to observe decays to proton-decaying levels in $^{21}$Na. Table \ref{table:21na-levels} lists the final levels in $^{21}$Na included within the present analysis. The $p_4$ and $p_5$ decays to proton-unbound states in $^{21}$Na are observed in the present experiment, though they cannot be resolved. However, the $E_\mathrm{x} = 2.797$-MeV state (populated by $p_4$ decays) in $^{21}$Na decays by $\gamma$-ray emission since the lifetime of the state ($\tau_{1/2} = 13(4) $ fs) is characteristic of a $\gamma$-ray lifetime. The $E_\mathrm{x} = 2.829$-MeV state is $J^\pi = 9/2^+$, requiring an $\ell_p = 4$ decay to the ground state of $^{20}$Ne. The single-particle Wigner limit for this decay is $\Gamma_p = 0.001$ eV, which is smaller than the expected $\gamma$-ray partial width, which is around $\Gamma_\gamma = 0.02$ eV for the decay of the mirror state in $^{21}$Ne. Practically, since the realistic width of a state is usually not more than around 10\% of the single-particle estimate, it is safe to conclude that the $\gamma$-ray decay dominates for this state for all realistic estimates of the resonance parameters of the state. Therefore, all of the states listed in Table \ref{table:21na-levels} decay by $\gamma$-ray emission.

Inspecting Fig. \ref{fig:BranchingRatios}, the sum of the proton branching ratios below $E_\mathrm{x} = 10$ MeV (which is the astrophysically relevant region, see Ref. \cite{PhysRevC.90.065806}) is typically exhausted by the observed proton decay channels. At higher energies, the branching ratio for decays to bound and $\gamma$-ray decaying unbound states can fall to around 60\%. This suggests that the $^{18}$Ne($\alpha,2p$)$^{20}$Ne reaction channel may be important at higher centre-of-mass energies, in agreement with the recent direct measurement with ANASEN \cite{PhysRevC.105.055806}, but that at lower centre-of-mass energies the contribution from $^{18}$Ne($\alpha,2p$)$^{20}$Ne is weaker. 

\section{Conclusions\label{sec:Conclusions}}

The $^{24}$Mg(p,t)$^{22}$Mg reaction has been measured with the K600 magnetic spectrometer at iThemba LABS, Cape Town, South Africa, with subsequent proton decays from the $^{22}$Mg recoils detected in an array of five double-sided silicon-strip detectors located around the target position of the K600. From the number of protons detected corresponding to decays to the $^{21}$Na ground and excited states, proton branching ratios were deduced.

The experimentally determined branching ratio for the p$_0$ emission is lower than the predictions from the statistical models TALYS \cite{TALYS}, NON-SMOKER \cite{NONSMOKER_web,RAUSCHER200147}, and SMARAGD \cite{2011IJMPE..20.1071R,SMARAGD}. Combining the measured p$_0$ branching with the assumption that the models correctly predict the total proton emission leads to a rescaling of the $^{18}$Ne($\alpha$,p$_0$)$^{21}$Na cross section which brings it into excellent agreement with the data by \cite{PhysRevLett.108.242701}. Under the same assumption the $^{18}$Ne($\alpha$,p)$^{21}$Na cross section remains unchanged. This is also in agreement with the recent data by \cite{2021arXiv210711498A}. 

While our experimental data led to an improvement in the prediction of the $^{18}$Ne($\alpha$,p$_0$)$^{21}$Na cross section, a better constraint on the astrophysical reaction rate for $^{18}$Ne($\alpha$,p)$^{21}$Na is not possible. The excellent reproduction of the $^{18}$Ne($\alpha$,p$_0$)$^{21}$Na data with the improved p$_0$ width, however, strengthens the case that this reaction can be described in the statistical Hauser-Feshbach reaction model.
\\

\begin{acknowledgements}
The authors thank the beam operators at iThemba LABS for the high-quality beam delivered for this experiment. It is a pleasure to thank the technical staff of the laboratory, particularly Mr Zaid Dyers, for their hard work in preparing the infrastructure for the \textsc{cake}. The \textsc{cake} was funded by the National Research Foundation (NRF) through the NEP grant 86052. JWB thanks the NRF for his postdoctoral fellowship and for the PDP funding of his PhD under which this study was completed. PA thanks the Trustees and staff of the Claude Leon Foundation for his postdoctoral fellowship, and especially thanks Maria Anastasiou for helpful discussions about direct measurements. TR is partially supported by the \textquoteleft ChETEC\textquoteright\ COST Action (CA16117). The authors would also like to thank Zach Meisel for his contribution and Kelly Chipps for helpful comments on a draft of this paper.  R.N. acknowledges financial support from the NRF through grant 85509.
\end{acknowledgements}

\bibliography{Mg22BranchingRatios}

\end{document}